\documentclass[showpacs,preprintnumbers, amsmath,amssymb,superscriptaddress,preprint]{revtex4}
\usepackage{graphicx}
\draft
\begin{document}
\newcommand {\be}{\begin{equation}}
\newcommand {\ee}{\end{equation}}
\newcommand {\bea}{\begin{eqnarray}}
\newcommand {\eea}{\end{eqnarray}}
\newcommand {\nn}{\nonumber}

\title{Panorama of Nodal Superconductors
}

\author{Hyekyung Won}
\author{Hyunja Jang}

\address{Department of Physics, Hallym University,
Chuncheon 200-702, South Korea
}

\author{David Parker}
\author{Stephan Haas}
\author{Kazumi Maki}

\address{Department of Physics and Astronomy, University of Southern 
California, Los Angeles, CA 90089-0484 USA}

\date{\today}

\begin{abstract}

Since 1979, many new classes of superconductors have been discovered,
including heavy-fermion compounds, organic conductors, high-T$_c$
cuprates, and Sr$_2$RuO$_4$.  Most of these superconductors are 
unconventional and/or nodal.  Therefore it is
of central importance to determine
the symmetry of the order parameter in each of these superconductors.  In 
particular, the angular-controlled thermal conductivity in the vortex state 
provides a unique means of investigating the nodal structure of the
superconducting energy gap when high-quality single crystals in the
extremely clean limit are available.  Using this method, Izawa et al have 
recently succeeded in identifying the energy gap symmetry of 
superconductivity in Sr$_2$RuO$_4$, CeCoIn$_5$, $\kappa$-(ET)$_2$Cu(NCS)$_2$, YNi$_2$B$_2$C, and PrOs$_4$Sb$_{12}$. 

\end{abstract}
\maketitle

\noindent{\it \bf 1. Introduction}

Since the appearance of anisotropic superconductors the determination
of their gap symmetries has been one of the central issues.\cite{volovik,sigrist}
For the high-T$_c$ cuprate superconductors angular-resolved photoemission 
spectroscopy (ARPES)\cite{shen, damascelli} and Josephson interferometry
\cite{tsuei, harlingen} provide a definitive signature of d$_{x^{2}-y^{2}}$-wave
superconductivity.  However, these methods have not yet been extended
beyond the high-T$_c$ cuprates.  In a remarkable paper Volovik\cite{volovik2}
showed that the quasiparticle density of states in nodal superconductors 
in a magnetic field is calculable within a semi-classical approximation.  Here
the Doppler shift\cite{maki} due to the supercurrent circulating around 
vortices plays a crucial role.  The resulting $\sqrt{H}$-dependence of
the specific heat in the vortex state has been observed experimentally in
YBCO\cite{moler,revas}, LSCO\cite{chen}, $\kappa$-(ET)$_2$ salts\cite{nakazawa},
and Sr$_2$RuO$_4$\cite{nishizaki,won}. Here $H$ is the magnetic field strength.

This approach has been extended in many directions.  K\"{u}bert and 
Hirschfeld\cite{kubert} considered the effect of non-zero temperature and
established the scaling law, as formulated by Simon and Lee\cite{simon}.  In 
particular, K\"{u}bert and Hirschfeld found the scaling function within the 
semi-classical approximation.  Won et al\cite{won} extended this approach for
the superfluid density, the spin susceptibility, and the nuclear spin lattice
relaxation rate.  Measurements of the thermal conductivity tensor in the vortex 
state provide an advantage, since the heat current direction provides
further directional information.  K\"{u}bert and Hirschfeld\cite{kubert} and 
Vehkter et al\cite{vehkter} have performed analyses in these directions.  
However, these authors did not take the spatial average over the vortex lattice
at the same time as the average over the Fermi surface, but instead assumed
a local thermal conductivity $\kappa_{ii}({\bf r})$.  It is known that the 
local quasiparticle
density of states mostly consists of bound states around vortex cores.
These bound states do not contribute to thermal conductivity when 
$H/H_{c2} \ll 1$.  On the other hand, the nodal excitations which do 
contribute to the thermal conductivity run through many unit cells of
the vortex lattice.  Furthermore Vehkter et al
chose a rather specific circular Fermi surface instead of the cylindrical Fermi 
surface commonly used in modeling the high-T$_c$ cuprates.
These problems were addressed in references 18-21. 
In these works simple expressions for the thermal conductivity are found  
in the superclean limit ($(\Gamma\Delta)^{1/2} \ll v\sqrt{eH} \ll \Delta(0)$, 
where $\Gamma$ is the quasiparticle scattering rate in the normal state and v
is the effective Fermi velocity).

Intuitively we understand that the Doppler shift generates quasiparticles in a
plane perpendicular to the vortex axis and the field direction.  When
this plane intersects the nodal directions, there is enhancement of the 
available quasiparticles, which we call nodal excitations.  Therefore, by 
changing the field direction one can sweep the Fermi surface with the plane
associated with the Doppler shift.  As this plane meets the nodal directions
both the specific heat and thermal conductivity are enhanced\cite{won2,dalim,won3}. 

In particular when the Fermi surface is quasi-2D or cylindrical, rotating the
magnetic field within the conducting plane yields valuable information
regarding the gap symmetry.  This was shown clearly in early experiments on
YBCO\cite{yu,aubin,ocana} and more recent experiments on Sr$_2$RuO$_4$\cite{izawa},
CeCoIn$_5$\cite{izawa2}, and $\kappa$-(ET)$_2$Cu(NCS)$_2$\cite{izawa3}.  
Such experiments, probing the angular dependence of the superconducting 
order parameter, 
indicate f-wave superconductivity for Sr$_2$RuO$_4$, d$_{x^{2}-y^{2}}$-wave
superconductivity for CeCoIn$_5$ and $\kappa$-ET$_2$Cu(NCS)$_2$, and s+g-wave superconductivity for
YNi$_2$B$_2$C and PrOs$_4$Sb$_{12}$.  These order parameters are shown in Fig. 1.  For $\Delta({\bf k})$ of PrOs$_4$Sb$_{12}$ see Fig. 10.

\begin{figure}[h]
\includegraphics[width=3cm]{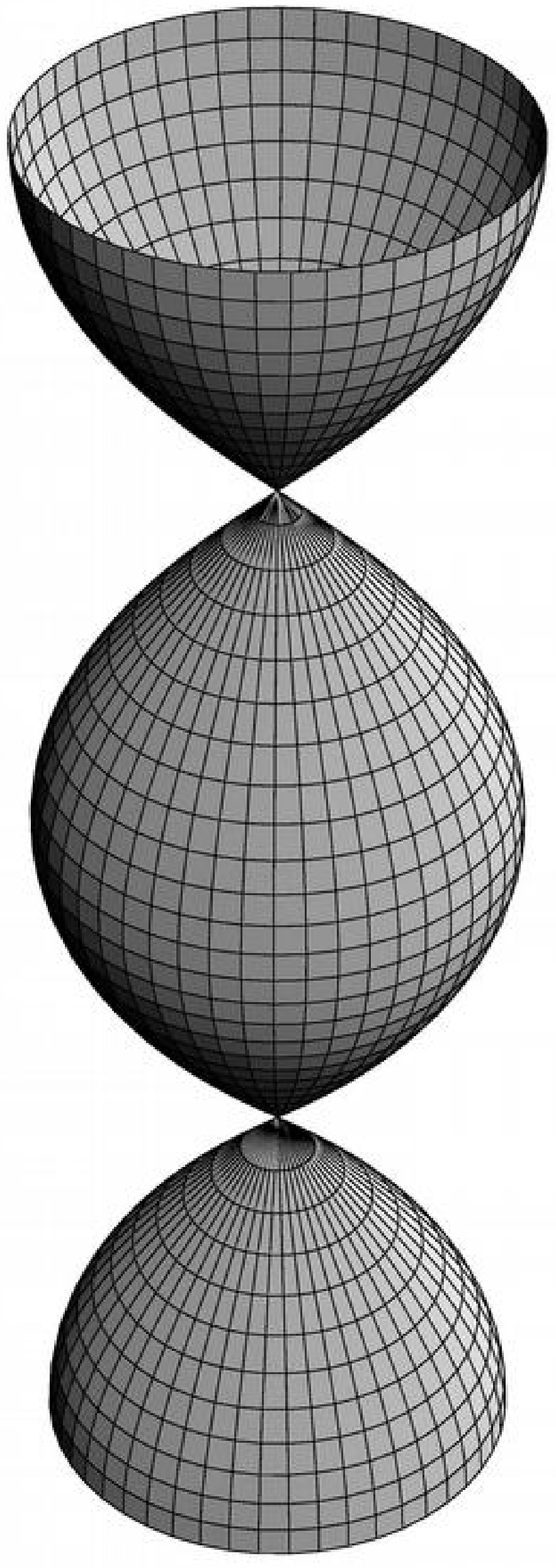}
\includegraphics[width=6cm]{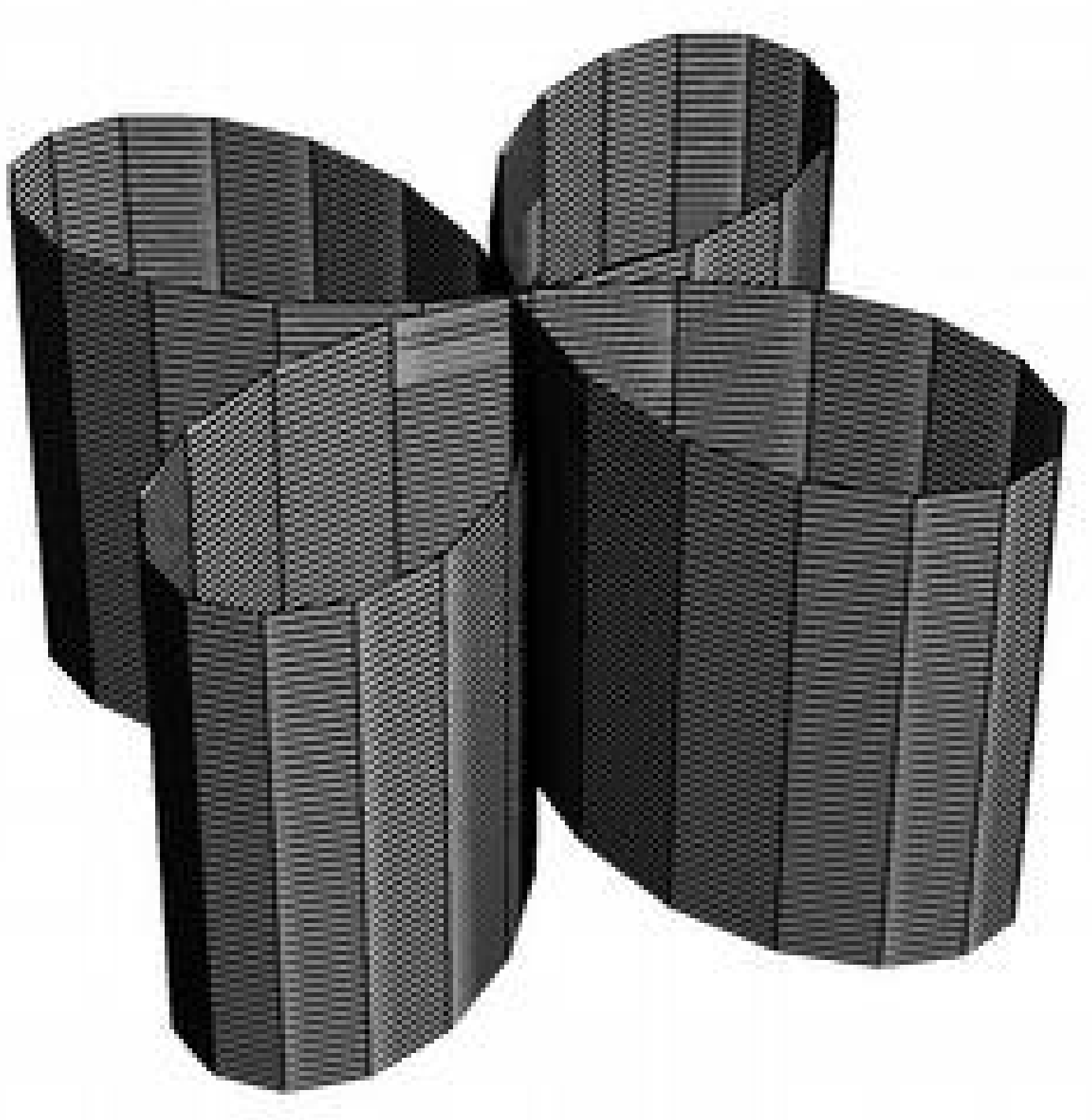}
\includegraphics[width=6cm]{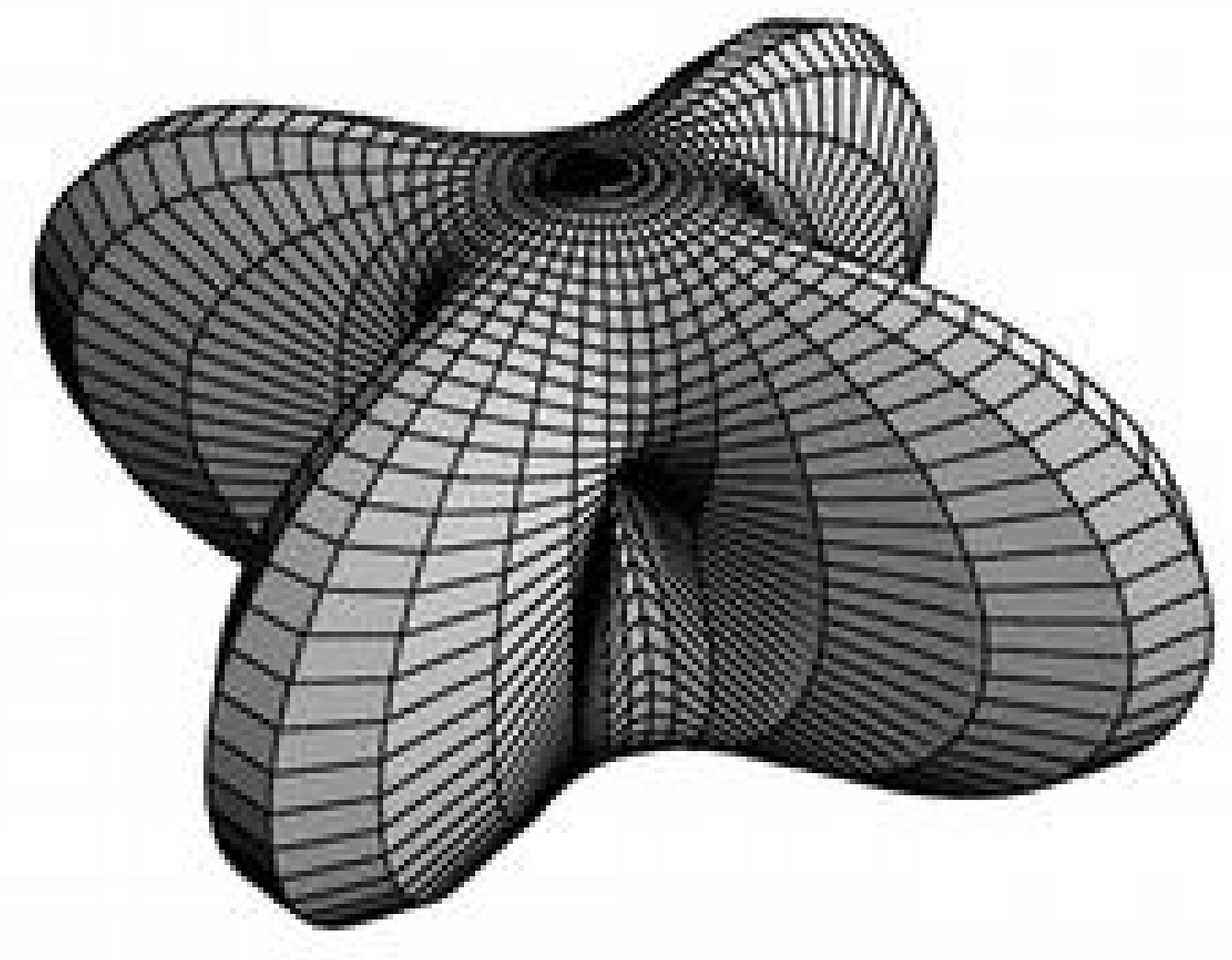}
\caption{
2D f-wave, $\rm d_{x^2 - y^2}$-wave, and
s+g-wave symmetry.
}
\end{figure}

In the following sections we focus on the salient features of 
superconductivity in high-T$_c$ cuprates, Sr$_2$RuO$_4$, $\kappa-$(ET)$_{2}$ salts, 
YNi$_2$B$_2$C, and PrOs$_4$Sb$_{12}$.

\noindent{\it \bf 2. High-T$_c$ Cuprate Superconductivity in a Nutshell}

The discovery of high-T$_c$ cuprate superconductivity in LBCO by Bednorz
and Muller\cite{bednorz} in 1986 took the superconductivity community by
surprise.  The subsequent excitement and confusion are well documented in
a textbook by Enz\cite{enz}.  In 1987 P.W. Anderson\cite{anderson} proposed
the 2-dimensional one-band Hubbard model and his famous dogma.  His central
theme is to understand the superconductivity in the presence of the
strong Coulomb repulsion.  In the meantime $d_{x^2-y^2}$-wave symmetry
of both the hole and the electron-doped high-T$_{c}$ cuprate superconductivity
has been established\cite{shen,damascelli,harlingen,tsuei}.  Furthermore, 
the mean-field theory (i.e. the generalized
BCS theory) for d-wave superconductivity works 
well\cite{won11,maki10,hussey}.  Also within the framework
of the BCS theory of d-wave superconductivity, May Chiao 
et al\cite{chiao1,chiao2} have derived
the crucial parameter $\Delta(0)/E_{F}$ of optimally doped YBCO and Bi-2212
through the measurement of thermal conductivity at $T < 1 K$.  Here $\Delta(0)$
is the maximum value of the energy gap at $T = 0 K$ and E$_F$ is the Fermi energy.
They found $\Delta(0)/E_{F} \simeq 1/14 $ and $1/10$ for YBCO and Bi-2212
respectively.  Regrettably there are no experimental data available
indicating the ratio $\Delta(0)/E_{F}$ in the underdoped and overdoped region
of YBCO and Bi-2212, though we have no reason to worry that this ratio
becomes substantially different.  First of all, these values tell us that high-T$_c$ cuprate
superconductivity is very far away from the Bose-Einstein (BE) condensate, but is
within the BCS regime.  The BE condensate 
clearly requires $\Delta(0) \sim E_{F}$.  Secondly, making use of Ginzburg's 
criterion the fluctuation effect in high-T$_c$ cuprate superconductivity
should be at most of the order of a few percent.

Thirdly, these $\Delta(0)/E_{F}$ values are incompatible with the assumption
that $\Delta(0) \simeq E_{F}$, which was made in solving the 
Bogoliubov-de Gennes equation in the
vortex state of d-wave superconductors\cite{franz,marinelli,ichioka10}.  In particular the approximation
$\Delta(0) \sim E_{F}$ appears to knock off all the bound states
around a single vortex in d-wave superconductors.  On the other hand, for 
$\Delta(0)/E_{F} = \frac{1}{10}$, one finds hundreds of bound states around
a single vortex\cite{kato1,kato2}.

In fact, the local density of states around a single vortex looks very
similar to the one found for s-wave superconductors\cite{gygi}.  
Second the approximation
$\Delta(0) \sim E_{F}$ appears to increase the $\sqrt{H}$ term in the
specific heat by a factor of 10 to 30\cite{ichioka10}.  Therefore we stress that if we
limit ourselves to the region $T \ll \langle |\bf{v}\cdot\bf{q}| \rangle 
\ll \Delta(0)$, the quasiclassical approach as discussed in Refs. 18 - 20 
provides the most reliable result so far available.

Unfortunately the approximation $\Delta(0) \simeq E_{F}$ is very popular
among the superconductivity community because it makes the calculation
much simpler\cite{ogata10,dahm10,miranovic}. 

As to the observation of these bound states around vortices in YBCO and
Bi2212, only a few bound states, if any, are 
observed\cite{maggio,hudson}.  But this appears 
to be due to the fact that the vortex core is filled with other order
parameters like antiferromagnetism\cite{lake10,kakuyanagi} 
or charge density wave.

In a popular paper Bob Laughlin\cite{laughlin10} proposed a very intuitive
picture of superconductivity in the presence of the Mott insulator.  We think
that another interesting question is how the superconductivity can 
survive in the presence of d-wave density wave\cite{benfatto,chakraverty,
dora10}.

In section 4 we shall discuss a very similar conflict between two order
parameters in the organic superconductor $\kappa$-(BEDT-TTF)$_{2}$X.

\noindent{\it \bf 3. Superconductivity in Sr$_2$RuO$_4$}

Superconductivity in Sr$_2$RuO$_4$ was discovered in 1994\cite{maeno}. 
The surprising prediction of triplet p-wave pairing and related chiral
symmetry breaking\cite{rice} was verified by muon spin rotation
experiments\cite{luke} and a flat Knight shift as seen by NMR 
measurements\cite{ishida}. Also, the triplet superconductivity implies 
clapping collective modes\cite{kee} and half-quantum vortices\cite{kee2}
as topological defects.  
Since sample quality has improved and single crystal Sr$_2$RuO$_4$
with T$_c$ $ \simeq$ 1.5 K have become available, experiments have found 
clearly nodal structures\cite{nishizaki,bonalde}.
This is inconsistent with the initially proposed fully
gapped p-wave model. Consequently, several f-wave models were
proposed\cite{haregawa,graf,won10}.

As shown in Ref. 19, most of these models are consistent with the 
specific heat data\cite{nishizaki} and the magnetic
penetration depth data\cite{bonalde}.  The p-wave model cannot account
for these measurements.  Furthermore, 
ultrasonic attenuation data by Lupien et al\cite{lupien} eliminates models
with vertical nodal lines parallel to k$_z$.  This leaves only the 2D f-wave
model with energy gap 
\bea
\Delta({\bf k}) = \Delta e^{\pm i\phi}cos(ck_z)               
\eea
where $e^{\pm i\phi} =  k_x \pm i k_y$ \cite{won4}.

Zhitomirsky and Rice\cite{zhitomirsky,annett} have 
proposed a multi-gap model where one of the superconducting order
parameters associated to the $\alpha$ and $\beta$ band has 
horizontal nodes at the Brillouin points\cite{mackenzie}.  However, 
this multi-gap model gives a 
two-fold-symmetric angular dependence
($\sim \cos(2\phi)$, where $\phi$ is the angle between the
heat current and the magnetic field) ten times larger than observed experimentally\cite{izawa}.
Moreover, the two-gap model cannot give universal heat conduction
as observed in $\kappa_{zz}$ \cite{suzuki}.  More recently, 
further tests of the 2D f-wave model for Sr$_2$RuO$_4$
were proposed\cite{dora,kee3,kee4}.
If the f-wave model is established for Sr$_2$RuO$_4$, it implies
that the 2D model used in the high-T$_c$ cuprates is inapplicable
to Sr$_2$RuO$_4$.  From the point of view of the electronic
interaction Sr$_2$RuO$_4$ should be a 3D system\cite{sato}.  Also
it is possible that p-wave pairing is forbidden in the electronic 
systems.  It appears we now have 3 f-wave superconductors: 
UPt$_3$\cite{heffner},
Sr$_2$RuO$_4$, and UNi$_2$Al$_3$\cite{ishida2}.  Finally, it is 
important to identify
the gap symmetry of the organic superconductor (TMTSF)$_2$PF$_6$ for which 
triplet pairing has already been established\cite{lee}.

Very recently Deguchi et al\cite{deguchi} have reported specific heat data for
0.12 K $\leq$ T $\leq$ 0.51 K and for magnetic field 0.05 T $\leq$ B
$\leq$ 1.7 T and rotating within the a-b plane.  One of the most 
surprising data is the fourfold term with cusps at $\phi = 0, \pi/2$, etc.,
which appears only for $B = 0.30 T, 0.60 T$ and $0.90 T$ at $T = 0.12 K.$
For data at $T=0.31 K$ the cusp feature becomes less clear.

Perhaps the most interesting question is 1) Does this include 2 energy
gaps?, and if so, 2) What is the gap symmetry of this new gap?

If we compare their Fig. 3 for $T=0.12K$, the angular dependence is very
similar to what one expects in s+g-wave superconductivity in the
presence of impurities\cite{maki20}.  In s+g-wave superconductors the impurity 
scattering induces a small energy gap.  Therefore, the most natural
interpretation of the above data is that the specific heat has
picked up four point-like mini-gaps located at $\theta = \pi/2$ and
$\phi=0,\pi/2$, etc.  The size of the minigap would be around 0.3 K.  
Then it is rather difficult to accommodate this with the Miyake-Narikiyo
model\cite{miyake}.  Also this quasi-nodal structure is difficult 
to accommodate with the f-wave order parameter above.  
Perhaps this experiment indicates
the presence of the second energy gap?  The gap function $\Delta({\bf k})$ with
minigap is shown in Fig. 2.  Here we take $|\Delta({\bf k})| \sim 
|\cos(\chi)|(1-1.8\cos(4\phi)\cos^{2}(\chi)+0.81\cos^{4}(\chi))^{1/2}$.

\begin{figure}[h]
\includegraphics[width=4.5cm]{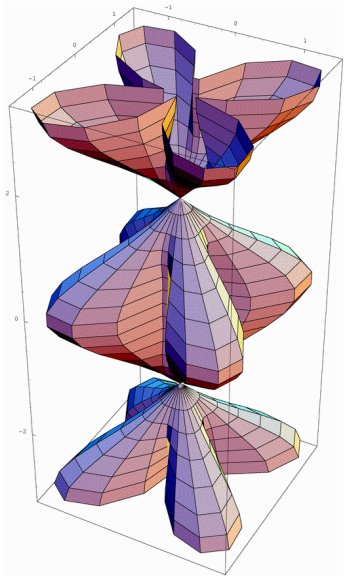}
\caption{
Proposed Gap Function for Sr$_{2}$RuO$_{4}$
}
\end{figure}

\noindent{\it \bf 4.  Gossamer Superconductivity in 
$\kappa$-(BEDT-TTF)$_{2}$X?}

The organic superconductors $\kappa$-(BEDT-TTF)$_{2}$X with
X = Cu[N(CN)$_2$]Br, Cu[N(CN)$_2$]Cl and Cu(NCS)$_2$ have the
highest superconducting transition temperature T$_c$ = 10-13 K
among organic conductors\cite{ishiguro}.  There are 
many parallels between high-T$_c$
cuprates and $\kappa$-(BEDT-TTF)$_{2}$X; the quasi-two dimensionality
of the Fermi surface, and the proximity to the antiferromagnetic
state.  More recently angular dependent STM\cite{arai} and  
angular dependent
thermal conductivity measurements in the vortex state\cite{izawa3} 
indicate d$_{x^{2}-y^{2}}$
superconductivity.  The nodal lines run\cite{izawa3} in the diagonal direction
of the b-c (i.e the conducting) plane.  Although d-wave superconductivity
has been speculated theoretically\cite{schmalian,louati}, the 
diagonal lines come as a surprise.
This indicates that perhaps the exchange of an antiparamagnon is not
adequate to generate d-wave superconductivity\cite{kuroki}.

Indeed from the thermal conductivity data from 
$\kappa$-(BEDT-TTF)$_{2}$Cu(NCS)$_4$, we can deduce that $\Delta({\bf k})=
\Delta(\cos(2\phi)-0.067)$.  This d+s-wave symmetry is somewhat similar
to that of YBCO\cite{won30}.

Perhaps more surprising is the sensitivity of the superconductivity to the
cooling rate, which has been studied by Pinteric et al\cite{pinteric} in
$\kappa$-(BEDT-TTF)$_{2}$Cu[N(CN)$_2$]Br.  To make the story simple
we consider two extreme cases only.  We define the relaxed sample as the 
one cooled very slowly down to 10 K.  For example this sample is left for
3 days in liquid N$_{2}$ and then cooled down for a fraction of a degree K 
per hour down to 10 K.  The other extreme is the quenched sample, wherein the
sample is dropped in liquid N$_2$, then cooled down to 10 K within a 
few hours.

Of interest, the superconducting transition temperature T$_c$ shows little 
dependence on the cooling procedure, with T$_c$ varying by only a few percent.
More surprising is the sensitivity to the cooling rate of the superfluid
density as measured by magnetic penetration depth.  The superfluid density
of the quenched sample is only 1-2 percent of the relaxed sample.  Also the 
temperature dependence of the relaxed sample exhibits a T-linear dependence
typical of d-wave superconductors\cite{carrington,pinteric2}.  
On the other hand, the superfluid density of the quenched
sample is very flat for T $\leq$ 0.2 T$_c$, which may be interpreted as a sign
of s-wave superconductivity.  Until recently there were debates on the symmetry
of superconductivity in $\kappa$-(BEDT-TTF)$_{2}$ salts: d-wave versus s-wave.
We therefore wonder if this controversy originates from the difference in
the cooling rate.  For example, Elsinger et al\cite{elsinger} did not describe
how their samples were cooled down.

Recently the effect of the cooling rate on the normal state of the three
$\kappa$-(BEDT-TTF)$_{2}$ salts has been reported\cite{muller2}.  These organic conductors
go through a glassy phase when they are cooled down from room temperature
to liquid N$_{2}$ temperature (70 - 90 K).

The BEDT-TTF molecules have ethylene groups attached to each end.  For
$T > 100 K$ these ethylenes are rotating freely.  As the temperature decreases
below 70 K these ethylenes cannot move freely but are settled in their 
equilibrium configurations.  Therefore it is very likely that in the 
relaxed samples these ethylene groups are relatively well ordered while in the 
quenched samples they are oriented randomly.

But it is not known at present how the randomness of the ethylene groups
interferes with the quality of the superconductivity.  From the insensitivity
of T$_c$ to the cooling procedure we infer that this cannot be the simple
effect of disorder.  Also Pinteric et al have shown the superconductivity
to be homogeneous.  More likely is that the disorder controls 
the appearance of another order parameter, a ``hidden order
parameter'', like unconventional density wave (UDW).  This forces 
the superconductivity to coexist with UDW.

As we have discussed in section 2 we call this type of superconductivity
``gossamer superconductivity''\cite{haas20}.  Unfortunately at present 
we do not know
what kind of order parameter is appropriate to characterize the glassy phase,
although we suspect it should be UCDW or USDW.  We believe that this is the
most fascinating question in $\kappa$-(BEDT-TTF)$_{2}$ salts.

\noindent{\it \bf 5. Superconductivity in YNi$_2$B$_{2}$C}

Superconductivity in YNi$_2$B$_2$C and LuNi$_2$B$_2$C was discovered
in 1994\cite{canfield}.  Recent interest has focused on these materials
because of their relatively high superconducting
transition temperatures of 15.5 K and 16.5 K, respectively.
There is a substantial s-wave component in the order parameter of
these compounds. This was shown by substituting Ni with a small amount of Pt
and observing the subsequent opening of the quasiparticle energy gap
as observed in specific heat measurements\cite{nohara}.  Nevertheless, recent
experiments clearly indicate the presence of nodal excitations\cite{zheng,izawa4}.

Furthermore, the upper critical field within the a-b plane exhibits
a clear four-fold symmetry\cite{methlasko, wang}.  The simplest
model to describe the superconductivity in YNi$_2$B$_2$C appears
to be\cite{maki2,izawa5,thalmeier}
\bea
 \Delta({\bf k}) = \frac{1}{2}\Delta (1-\sin^{4}\theta \cos(4\phi))         
\eea
or s+g-wave superconductivity.  Here $\theta$ and $\phi$ are the polar and azimuthal
angles, respectively, describing the direction of ${\bf k}$.  The 
corresponding $\Delta(\bf {k})$ is shown in Fig. 1(c).
The quasiparticle density of states is given by 

\bea
G(E) \equiv N(E)/N_{0} = |E| Re\langle \frac {1}{(E^{2}-\Delta^{2}({\bf k}))^{1/2}}\rangle
\eea
where $\langle---\rangle$ means $(1/4\pi)\int d\Omega---$. This is shown 
in Fig. 3.  In particular, for $|E|/\Delta \ll 1$, we obtain
\bea
G(E)= \frac{\pi|E|}{4\Delta} + \frac{9}{16}\left(E/\Delta\right)^2 + \ldots     
\eea

\begin{figure}[h]
\includegraphics[width=7cm]{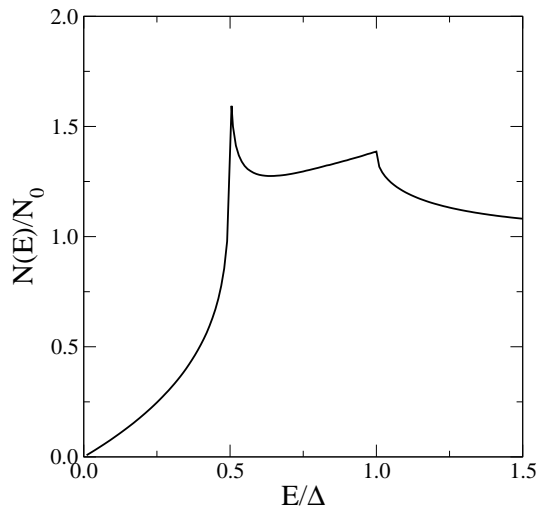}
\caption{
Quasiparticle density of states of an s+g-wave superconductor.
}
\end{figure}

This gives rise to power laws in the specific heat and other quantities
as follows\cite{jang}:

\bea
\frac{C_s}{\gamma_N T}& =& \frac{27\zeta(3)}{4\pi)}(T/\Delta) + 
\frac{63}{50}(T/\Delta)^{2} + \ldots  \\      
\frac{\chi_{s}(T)}{\chi_N}& =& \frac{\pi \ln(2)}{2}(T/\Delta) + 
\frac{3\pi^{2}}{16}(T/\Delta)^{2}  + \ldots \\      
\frac{\rho_{sab}(T)}{\rho_{sab}(0)}& =& 1 - \frac{3\pi\ln{2}}{4}(T/\Delta) - 
\frac{5\pi^{2}}{32}(T/\Delta)^{2} + \ldots  \\   
\frac{\rho_{sc}(T)}{\rho_{sc}(0)}& =& 1 - \frac{\pi^{2}}{4}(T/\Delta)^{2}- 
\frac{783\pi}{256}(T/\Delta)^{3} + \ldots    \\ 
T_{1}^{-1}/T_{1N}^{-1}& \simeq & \frac{\pi^{4}}{48}(T/\Delta)^{2} +
\frac{81\pi\zeta(3)}{32}(T/\Delta)^{3} + \ldots 
\eea
where $\gamma_N$ is the Sommerfeld coefficient, and $\zeta(3) \simeq$  1.202
is the Riemann zeta function.  We note that the presence of point nodes
in the a-b plane creates an anisotropic temperature
dependence in the superfluid density.

These specific heats, spin susceptibility and anisotropic superfluid 
densities are compared with those
of a d-wave superconductor in Figs. 4, 5, and 6(a) and 6(b).  
\begin{figure}[h]
\includegraphics[width=7cm]{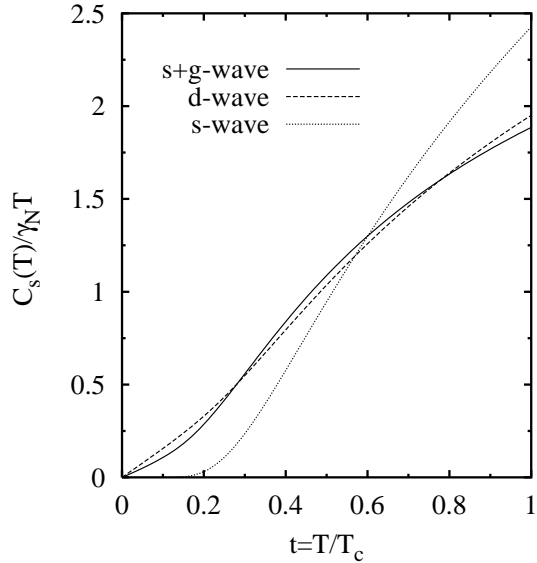}
\caption{Specific heats normalized by normal state specific heat are shown
for s+g-wave,s-wave and d-wave superconductors. 
$\gamma_{N}$ is the Sommerfeld constant.
}
\end{figure}

\begin{figure}[h]
\includegraphics[width=7cm]{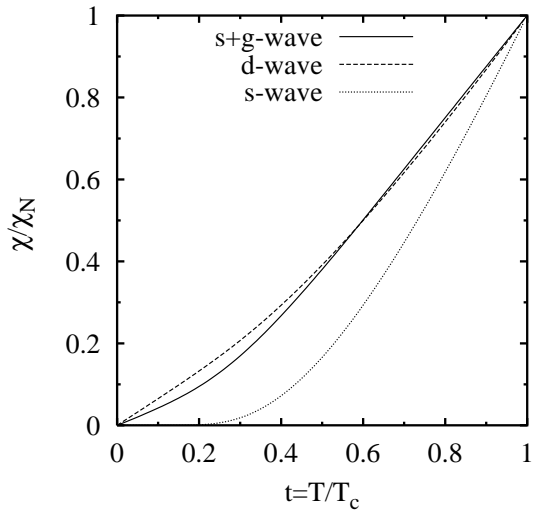}
\caption{The spin susceptibility normalized by that of the normal state
are shown for s+g-wave, s-wave and d-wave superconductors.}
\end{figure}

\begin{figure}[h]
\includegraphics[width=7cm]{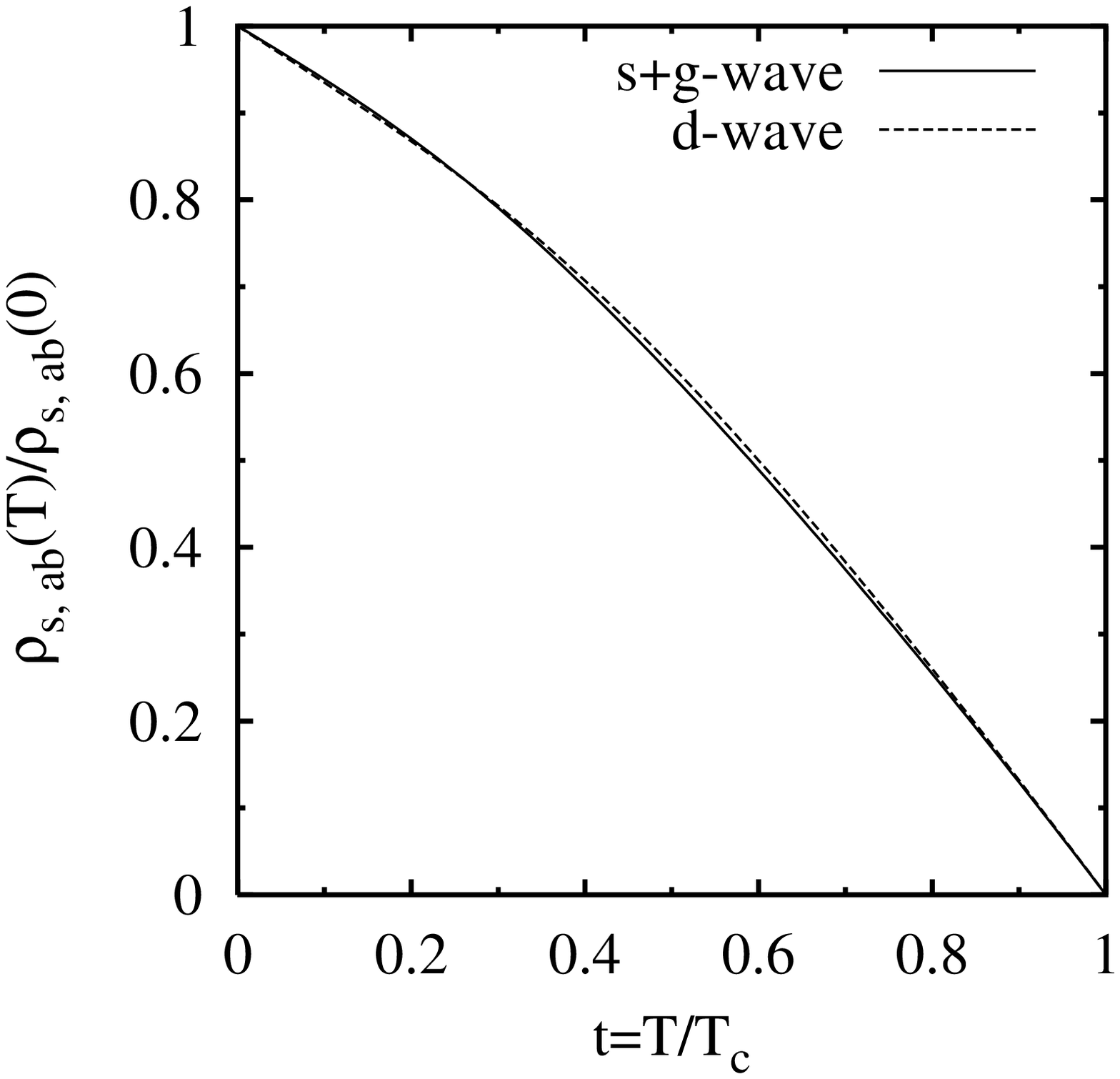}
\includegraphics[width=7cm]{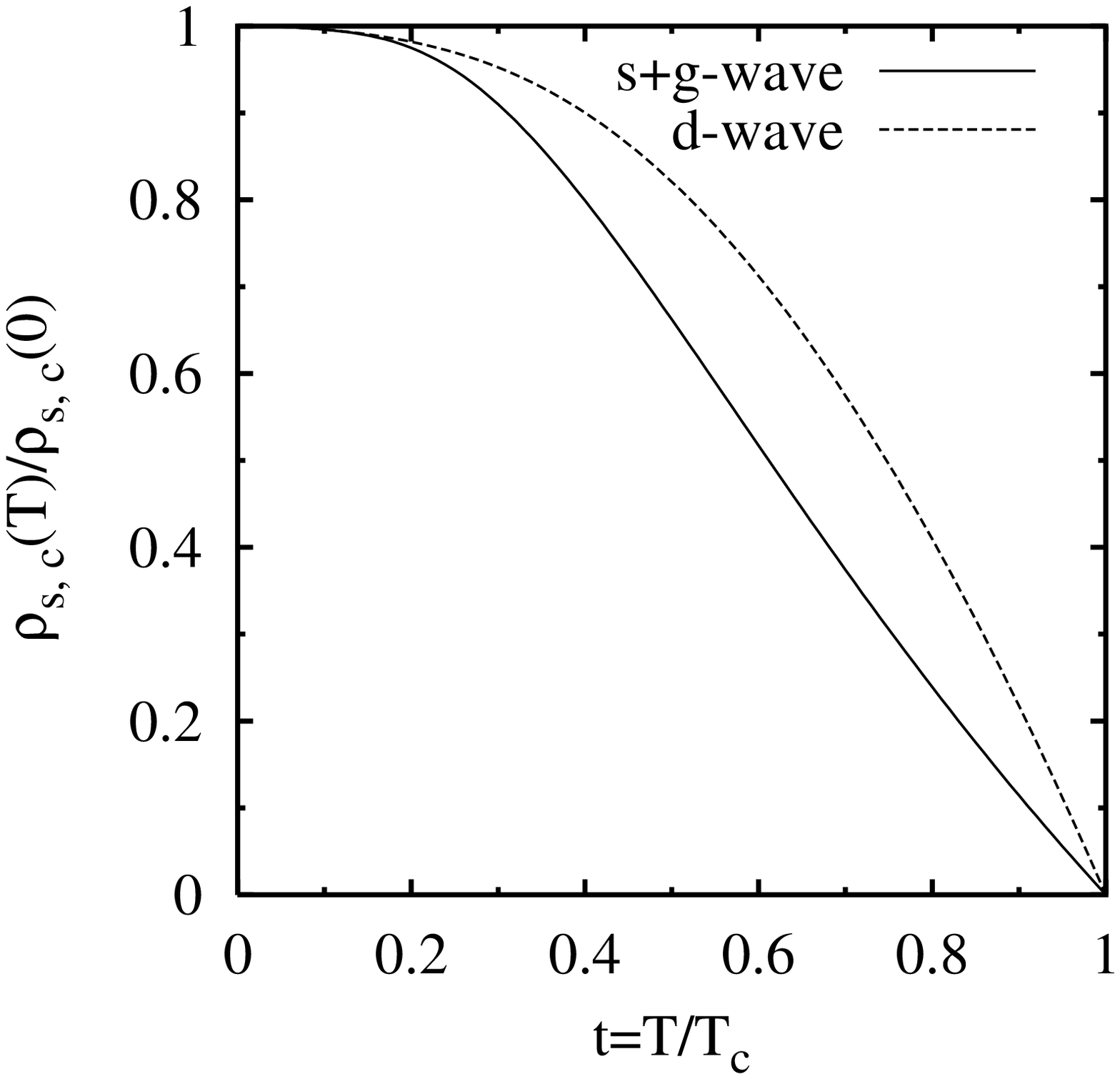}
\caption{
Superfluid density in the ab-plane, and parallel to
the c-axis, for s+g and d-wave superconductors}
\end{figure}

In the vortex state and in the superclean limit $(\Gamma 
\ll \tilde{v}\sqrt{eH}$, where $\tilde{v} = \sqrt{v_{a} v_{b}}$ and $\Gamma$
is the scattering rate in the normal state) the specific heat and 
other quantities are given by

\bea
\frac{C_s}{\gamma_N T}& =& 
\tilde{v}\frac{\sqrt{e H}}{2\Delta}I(\theta, \phi) \\   
\frac{\chi_{s}(T)}{\chi_N}& =& 
\tilde{v}\frac{\sqrt{e H}}{2\Delta} I(\theta, \phi) \\   
\frac{\rho_{sab}(T)}{\rho_{sab}(0)} & =& 
1 - \frac{3\tilde{v}\sqrt{eH}}{4\Delta}I(\theta, \phi) \\  
\kappa_{zz}/\kappa_n & = &\frac{x}{4ln(2/x)}    
\eea
where
\bea
x = \frac{2\tilde{v}\sqrt{eH} I(\theta, \phi)}{\pi\Delta}
\eea
and
\bea I(\theta,\phi) = \frac{1}{2} ((1-\sin^{2}\theta \sin^{2}\phi)^{1/2} +
(1 - \sin^{2}\theta \cos^{2}\phi)^{1/2})  
\eea

Here $\theta$ and $\phi$ are the polar and azimuthal angles
describing the direction of the magnetic field.

Recent thermal conductivity\cite{izawa5} and 
specific heat data \cite{park} establish experimentally this
striking angular dependence.  In Fig. 7 we show the experimental data
together with the theoretical expression.  The cusps at $\phi = 0, \pi/2$,
etc. for $\theta = \pi/2$ clearly indicate the presence of point 
nodes.  In the presence of line nodes the angular dependence of
$\kappa_{zz}$ is well approximated by $\cos(4\phi)$.  As is seen 
from the denominator of Eq. 13,  the effect of impurity scattering is
very unusual\cite{lee.s,yuan}.

\begin{figure}[h]
\includegraphics[width=15cm]{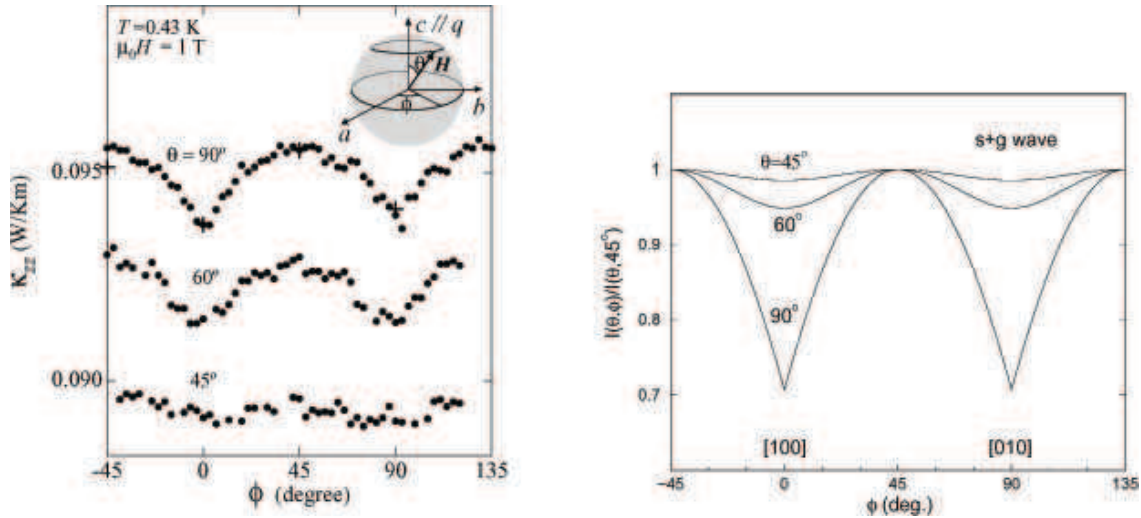}
\caption{Experimental and theoretical angular dependence of $I(\theta,\phi)$}
\end{figure}

Regarding Raman scattering, we present here results from a concurrent
work\cite{jang} wherein theoretical calculations, based on the s+g
symmetry, for the modes $A_{1g}$,
$B_{1g}$ and $B{2g}$ are performed.  The comparison between theory and
experiment\cite{euro} is shown in Figure 8, with the theoretical results on the left    
and the experimental results, taken at 6 K, on the right.  Strong agreement
is seen for the whole energy range in the $A_{1g}$ mode, with fair agreement 
for the other two modes.  Notably absent from the $B_{2g}$ data is
the secondary
cusp feature found at $\omega = 2\Delta$; we believe that this may be
a temperature-related effect and eagerly await the results of
lower-temperature experiments.

From the experimental data\cite{euro} we can extract $\Delta(0)$=50.4 K
and 64.7 K for YNi$_{2}$B$_{2}$C and LuNi$_{2}$B$_{2}$C respectively.  
On the other hand, the weak coupling theory gives $\Delta(0)$ = 
42.2 K and 43.3 K respectively, where we need $\Delta(0)/T_{c}$ = 2.72.
Therefore we may conclude that YNi$_{2}$B$_{2}$C is close to the weak 
coupling limit whereas LuNi$_{2}$B$_{2}$C is in the moderately strong
coupling limit.

\begin{figure}[h] 
\includegraphics[width=10cm,angle=270]{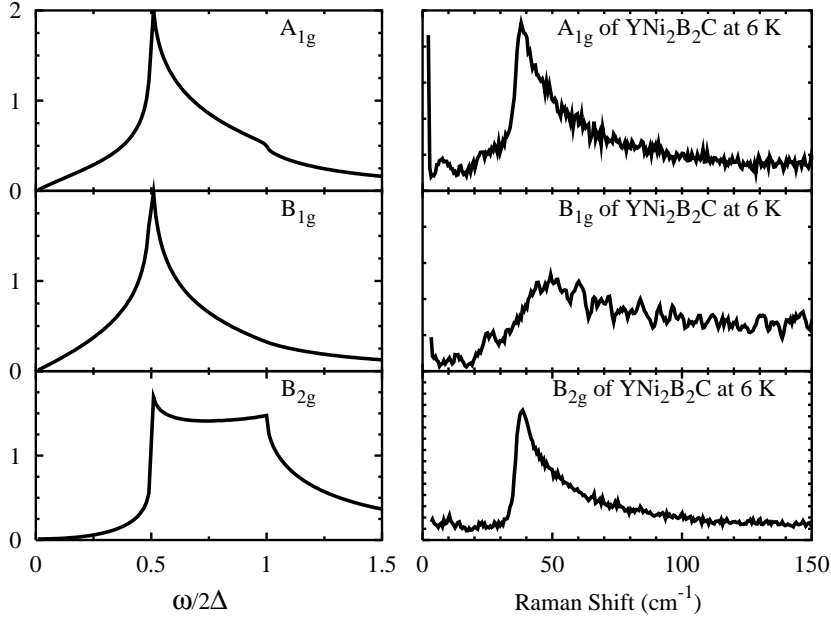}
\caption{Theoretical Raman spectra for the $A_{1g},B_{1g}$
and $B_{2g}$ modes
using the s+g model at $T = 0 K$ (left side)
and experimental Raman spectra
for YNi$_{2}$B$_{2}$C taken at $T = 6 K$ (right side) are shown.}  
\end{figure}

In the absence of a magnetic field the quasiparticle spectrum
in the presence of impurities is determined by\cite{abrikosov}
\bea
\tilde{\omega} = \omega +
\Gamma\tilde{\omega}
\langle\frac{1}{\sqrt{(\tilde{\Delta}-\Delta f/2)^{2} - 
(\tilde{\omega})^{2}}}\rangle 
\eea
and  
\bea \tilde{\Delta} = \Delta/2 + \Gamma \langle\frac{\tilde{\Delta} - 
\Delta f/2}{\sqrt{(\tilde{\Delta}-\Delta f/2)^{2} - 
(\tilde{\omega})^{2}}}\rangle 
\eea
where $\tilde{\omega}$ and $\tilde{\Delta}$ are the renormalized Matsubara frequencies
and order parameter, respectively, and $\langle --- \rangle$ means 
$1/(4\pi) \int d\Omega ---$.

Then the quasiparticle density of states is given by
\bea
G(E) = N(E)/N_0 = |E|\left(Re\langle\frac{1}{\sqrt{\tilde{\omega}^{2} - 
(\tilde{\Delta} - \Delta f/2)^{2}}}\rangle\right) 
\eea
where $G(E)$ is evaluated at $\omega = E$.
The DOS for a few $\Gamma$ are shown in Fig. 8.
Here we have used the so-called Born limit due to the 
presence of a substantial s-wave component in $\Delta({\bf k})$.
The unitary limit gives virtually the same result as the
Born limit.
\begin{figure}[h]
\includegraphics[width=8cm]{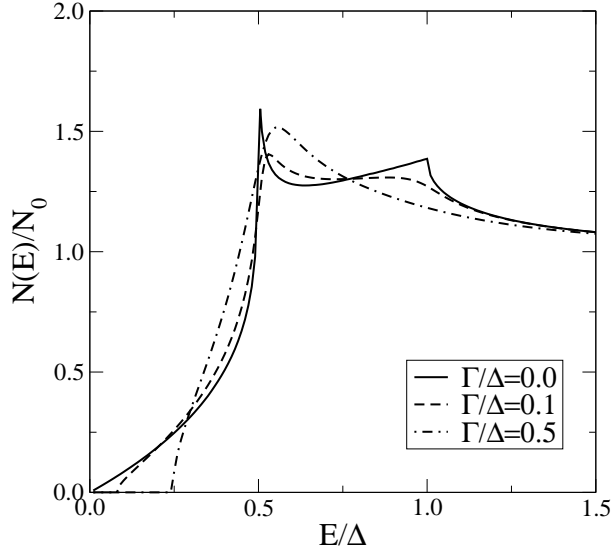}
\caption{Density of states for various impurity scattering levels}
\end{figure} 
The most unusual feature one notices is the immediate appearance
of an energy gap when $\Gamma \neq 0$.  This is in contrast to
the usual nodal superconductors with line nodes\cite{sun}.  In
particular, the energy gap is given by $\omega_g = \Gamma (1 + 2\Gamma/\Delta)^{-1}$.
This has a number of consequences\cite{won4}.  First of all, in the absence of
a magnetic field both the specific heat and the thermal
conductivity vanish exponentially.
\bea
C_s/T, \kappa_{ii}/T \sim (\omega_{g}/T)^{3/2}e^{-\beta \omega_g}
\eea

There is no universal heat conduction unlike other nodal 
superconductors\cite{sun,leep.a}.  This remarkable effect
of impurity scattering on nodal excitation is clearly shown by
Kamata\cite{kamata}.  In Fig. 9 we show the thermal 
conductivity data of Y(Ni$_{1-x}$Pt$_x$)$_2$B$_2$C with x=0.05.
As is readily seen the fourfold term typical of the pure
YNi$_2$B$_2$C has vanished completely.
\begin{figure}[h]
\includegraphics[width=8cm]{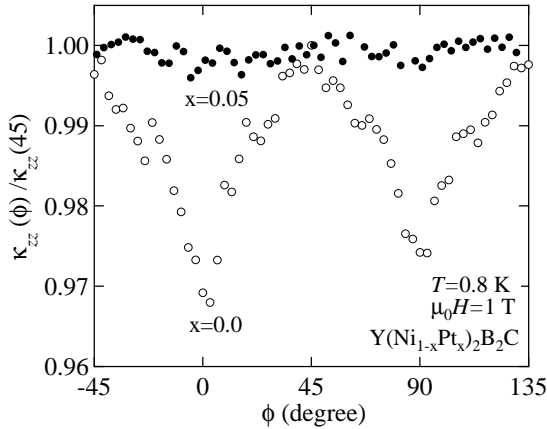}
\caption{Thermal conductivity data for Y(Ni$_{1-x}$Pt$_{x}$)$_{2}$B$_{2}$C
with x=0.05.}
\end{figure}
For example, the quasiparticle density of states in the 
presence of both a magnetic field and impurities is given
by\cite{maki20}
\bea
G({\bf{H}}, \Gamma)& = & x\cos^{-1}(y)\theta(1-y)               
\eea
where x is defined in Eq.(14) and $y = \frac{\Gamma}{\Delta x}$.

Further we obtain
\bea
\frac{\kappa_{zz}}{\kappa_n}& =& \frac{x}{2\cosh^{-1}(1/x)} ((1-y^{2})^{3/2} - 
\frac{3y}{2} (\cos^{-1}(y) - y\sqrt{1-y^{2}}))\theta(1-y) \\    
\frac{\kappa_{xx}}{\kappa_n}& =& \frac{3}{2\cosh^{-1}(1/x)}(\frac{x'}{x})^{2}(\cos^{-1}(y') - 
y'\sqrt{1-y'^{2}})\theta(1-y')
\eea
where
\bea
x' = \frac{1}{\pi}\frac{\tilde{v}\sqrt{eH}}{\Delta}(1 - \sin^{2}\theta \cos^{2}\phi)^{1/2}  ,  
y' = \frac{\Gamma}{\Delta x'}.
\eea

These expressions indicate clearly that the nodal excitations
are eliminated when $y > \sqrt{2}$.

\noindent{\it \bf 6. Puzzle of Superconductivity in PrOs$_4$Sb$_{12}$}

New heavy-fermion superconductivity in the skutterudite PrOs$_4$Sb$_{12}$
with T$_c$ = 1.8 K was discovered quite recently\cite{bauer}.
The existence of two distinct superconducting phases and the 
presence of point nodes in $\Delta({\bf k})$ are of great 
interest\cite{vollmer,kotegawa}.  Indeed, recent angular-dependent
magnetothermal conductivity data\cite{izawa6,maki3} indicate the
presence of two phases, with 4 point nodes in the ab-plane in 
phase A and 2 point nodes parallel to the b axis in phase B (see 
Fig. 10).

\begin{figure}[h]
\includegraphics[width=5cm]{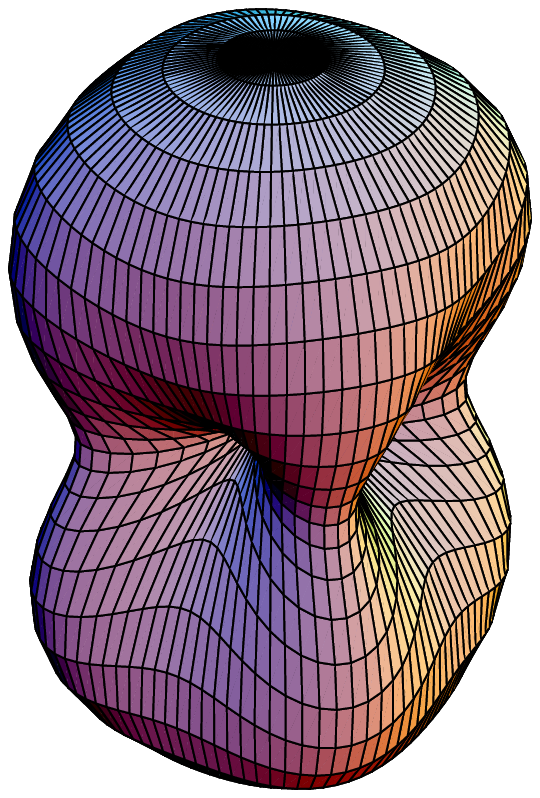}
\includegraphics[width=5cm]{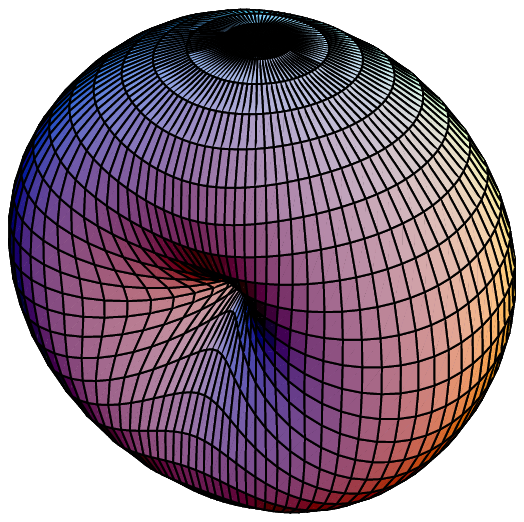}
\caption{Gap functions for A and B phase (right side) of PrOs$_4$Sb$_{12}$}
\end{figure} 

In order to accommodate the observed nodal structure within the
cubic symmetric crystal, the following order parameters are
proposed:
\bea
\Delta_{A}({\bf k})& =& \Delta(1-k_{x}^{4}-k_{y}^{4}) \\ 
\Delta_{B}({\bf k})& =& \Delta(1-k_{y}^{4}) 
\eea 
for the A and B phases, respectively.  

From an experimental point of view, it must be acknowledged that whether
the superconductivity is the spin singlet or the spin triplet 
is unclear.  Here we have assumed the spin singlet symmetry.
However, a muon spin rotation experiment indicates the
presence of remanent magnetization which may be
indicative of the spin triplet symmetry\cite{aoki}.  
Indeed alternative models have been proposed\cite{goryo,ichioka,chia},
although these models cannot describe the observed angular 
dependence of the thermal conductivity.

The models given in Eq.(24) and (25) have the $T^{2}$-specific
heat
\bea
\frac{C_s}{\gamma_N T} = \left\{ \begin{array}{ll} 
27\zeta(3)(T/\Delta)/4\pi & \mbox{for A phase}  \\
27\zeta(3)(T/\Delta)/8\pi & \mbox{for B phase} \end{array} \right. 
\eea

Of more interest is the 
anisotropic superfluid density\cite{maki3}.  In particular, in the 
B-phase
\bea
\frac{\rho_{s\parallel}(T)}{\rho_{s\parallel}(0)} = 1 - 
\frac{3\pi \ln{2}}{4}\frac{T}{\Delta}
- \frac{\pi^{2}}{16}\left(\frac{T}{\Delta}\right)^{2} + \ldots  \\      
\eea
and 
\bea \frac{\rho_{s\perp}(T)}{\rho_{s\perp}(0)} = 1 - 
\frac{\pi^{2}}{16}\left(\frac{T}{\Delta}\right)^{2} - 
\frac{21\pi\zeta(3)}{128}\left(\frac{T}{\Delta}\right)^{3} +   \ldots        
\eea
where the suffixes $\parallel$ and $\perp$ indicate directions
parallel and perpendicular to the nodal directions.

Very recently surprising superfluid density measurements of
PrOs$_4$Sb$_{12}$ were reported\cite{chia}.  Chia et al
measured the superfluid density parallel to each of the three crystal
axes and found that the superfluid densities are isotropic and 
decrease like T$^{2}$ at low temperatures.  Both the isotropy
and the T$^{2}$ dependence are incompatible with all models
for PrOs$_4$Sb$_{12}$ proposed to date. 
However, if one assumes that the nodal points in the B-phase
are always aligned parallel to ${\bf H}$ when the samples
are cooled in a fixed magnetic field, our model for the B-phase
fits the experimental data precisely.
This is plausible, since the magnetic field is the only
symmetry-breaking parameter in the present model.  Also, a simple
analysis of the effect of the magnetic field on the superconducting
order parameter tells us that the effect of the magnetic field
is minimized when the field is parallel to the nodal directions.

In the presence of a magnetic field the specific heat and
the thermal conductivity in the superclean limit are given 
by\cite{maki3}
\bea  
\frac{C_s}{\gamma_N T}& =& \pi x_{i}/4  \\
\frac{\kappa_{zz}^{A}}{\kappa_n}& =& 
\frac{x_{A}}{2 ln(2/x_A)} \frac{1+\pi x_{A}/2 +
 31 x_{A}^{2}/40}{1 + 31 x_{A}^{2}/64}  \\
\frac{\kappa_{zz}^{B}}{\kappa_n}& =&  
\frac{x_{B}}{2 ln(2/x_B)} \frac{1+\pi x_{B}/12 +
31 x_{B}^{2}/40}{1 + 31 x_{B}^{2}/64}   
\eea
where 
\bea x_{A} = \frac{v\sqrt{eH}}{\pi \Delta}
((1-\sin^{2}(\theta)\cos^{2}(\phi))^{1/2} +
(1-\sin^{2}(\theta)\sin^{2}(\phi))^{1/2})
\eea
and
\bea x_{B} = \frac{v\sqrt{eH}}{2\pi\Delta}
(1-\sin^{2}(\theta)\sin^{2}(\phi))^{1/2}
\eea

Here we assumed that $\Delta({\bf k})$ is given by equations
(24) and (25) for the A-phase and B-phase, respectively.  The
above expressions are consistent with the thermal
conductivity data\cite{izawa6}, although experimental data at lower
temperatures would help to further clarify this point.

Therefore the model proposed in Refs. 112 and 113 appears to be the
most consistent with the available data.  On the other hand
we have assumed the spin singlet pairing, while some experiments
appear to indicate spin triplet pairing\cite{aoki}.  Further
clarification on this question is highly desirable.

\noindent{\it \bf 7. Outlook}

In the past decade we have witnessed the identification of
the gap symmetry of the high-T$_c$ cuprates through 
ARPES\cite{shen, damascelli} and phase sensitive Josephson
interferometry\cite{harlingen,tsuei}.  However, these
techniques do not appear to be practical to use on heavy-fermion
superconductors or organic superconductors.

In recent years measurements of
the angular dependent magnetothermal conductivity have provided
a unique alternative means to explore the gap symmetry of superconductors
with T$_c$ $\sim$ 1-2 K.  In this way the gap symmetries of
Sr$_2$RuO$_4$, CeCoIn$_5$, $\kappa - (ET)_{2} Cu(NCS)_{2}$, YNi$_2$B$_2$C
and PrOs$_4$Sb$_{12}$ have been identified.  The next
step will be to interpret these gap symmetries in terms of
available interaction terms in these systems.

In this journey we have discovered that superconductivity with
mixed representations play a crucial role in both YNi$_2$B$_2$C
and PrOs$_4$Sb$_{12}$.  This is very surprising, but it appears that
one must accept this new development.  Clearly this will
open up a new vista in the rich field of nodal superconductors.

{\bf Acknowledgements}

We would like to express our thanks for useful collaborations
and discussions with Thomas Dahm,Peter Thalmeier, Hae-Young Kee, Balazs
Dora, Attila Virosztek, Konomi Kamata, Koichi
Izawa,Yuji Matsuda and Silvia Tomi\'{c} on related subjects.  
Also one of us (KM) thanks the hospitality and support of 
the Max-Planck Institute for the Physics of Complex Systems 
at Dresden, where part of this work was done.

\end{document}